\documentstyle[12pt]{article}
\begin{document}
\title{\large \hspace{10cm} ITEP-26/97 \\ \hspace{10cm} May 1997 \\
\vspace{1cm}
\LARGE \bf High orders of the perturbation theory for hydrogen atom in
magnetic field}
\author {V.A.GANI\thanks{E-mail addresses:
gani@vitep5.itep.ru, Vahid.Gani@itep.ru, gani@heron.itep.ru}{\,}
\\
{\it Moscow State Engineering Physics Institute (Technical University),}\\
{\it Kashirskoe shosse, 31, Moscow, 115409, Russia}\\
{\it and}\\
{\it Institute of Theoretical and Experimental Physics, Russia}\\
\\
A.E.KUDRYAVTSEV \thanks{E-mail address: kudryavtsev@vitep5.itep.ru} and
V.M.WEINBERG \thanks{E-mail address: wein@vitep5.itep.ru}\\
\\
{\it Institute of Theoretical and Experimental Physics,}\\
{\it B.Cheremushkinskaja, 25, Moscow, 117259, Russia}\\
}
\date{}
\maketitle
\vspace{1mm}
\centerline{\bf {Abstract}}
\vspace{3mm}
The states of hydrogen atom with principal quantum number $n\le3$ and zero
magnetic quantum number in constant homogeneous magnetic field ${\cal H}$ are
considered. The coefficients of energy eigenvalues expansion up to 75th order
in powers of ${\cal H}^2$ are obtained for these states. The series for energy
eigenvalues and wave functions are summed up to ${\cal H}$ values of the order
of atomic magnetic field. The calculations are based on generalization of the
moment method, which may be used in other cases of the hydrogen atom
perturbation by a polynomial in coordinates potential.

\newpage

\begin{center}
\bf
1. INTRODUCTION
\end{center}

\bigskip
   A new aspect of the problem of hydrogen atom in constant
electric (${\cal E}$) and magnetic (${\cal H}$) external fields was
observed recently. It was found [1], that the asymptotic of the
perturbation series in powers of ${\cal E}$ considerably changes at
some ${\cal H}$ values. This change is related with complex solutions
of classical equations of motion, previously not taken into account.
A look at this phenomenon from the expansion in powers of ${\cal H}$
point of view could be useful. Our work is aimed just in this direction.
An effective method for building the perturbation series is offered and
the asymptotic of this series for the Zeeman effect is discussed here.

   The moment method for high orders of the perturbation theory
evaluation, possessing additional possibilities as compared with the other
known recurrent methods, was introduced in Ref. [2].
It was described as an expedient for dimensional expansion
investigation --- in a problem leading to effective isotropic
anharmonic oscillator. Then it was applied to dimensional expansion
for three body problem [3], where the effective anharmonic oscillator
is anisotropic.

   We turn our attention to the fact, that the perturbation of the
hydrogen atom by a potential of polynomial form is also convenient
to investigate by the moment method. The constant homogeneous
electric and magnetic fields are referred just to this kind of
perturbations. The advantages of the moment method are clearly seen
in the Zeeman effect problem. High orders of the perturbation theory
(PT) using different approaches were studied in Refs. [4-6].
The method, based on the group theory was applied.
36 coefficients of the hydrogen's ground state energy expansion in powers of
${\cal H}^2$ were published in Ref. [6]. For excited states Zeeman's PT
coefficients not higher than of the third order in ${\cal H}^2$ are
given in Refs. [7-10]. The variables in the Schr\"odinger
equation for the Zeeman effect can not be separated and it makes
computation of higher PT orders more difficult. The moment method
does not require variables separation. Besides, as it will be shown
here, this method can be applied to degenerate states. Obviously this
possibility is essential for most of the hydrogen atom states.

   Apparently the logarithmic perturbation theory (LPT) was most
frequently used for recurrent evaluation of PT coefficients up to now
[11-15]. Owing to its simple algebraic structure, LPT  allowed to
compute the highest orders corrections for some cases. Thus, for the
Stark shift of the hydrogen ground state 160 orders of PT were
obtained this way [16]. But LPT is not free from restrictions. Even
one node of the wave function leads to considerable more complicated
computation procedure [17]. Much more complicated LPT looks in
problems, where variables can not be separated. Without variables
separation only a few initial PT orders were computed with the help
of LPT for the hydrogen atom in electric and magnetic fields [9,10,14].

   It is also worth mentioning here the old PT version, based on the
generalized virial and Hellmann--Feynman theorems, which anticipated
the modern moment method. In Refs. [18,19] recurrence relations
were written for problems with spherical symmetry, allowing
to compute energy  eigenvalues corrections for states with nodes in
as simple way as for nodeless states. But the moments introduced in
these papers were diagonal ones and that is why the region of
applicability of the method is restricted. For instance, in this formalism
it is impossible to obtain the corresponding wave function.

   The recurrence relations of the Ader moment method will be obtained
below. Two examples illustrate their applications: one is referred
to non-degenerate state and the other to the case of degeneracy.
Then the asymptotic of the numerically obtained energy expansion
coefficients is considered and the results of summation of perturbation
series are given. It will be shown on example of the ground state, how
the wave function of the perturbed hydrogen atom can be obtained within
the moment method.

\bigskip

\begin{center}
\bf
2. MAIN RECURRENCE RELATION
\end{center}

\bigskip

  Consider the state of the hydrogen atom $|\psi_0\rangle=|n,l\rangle$
with principial quantum number $n$, angular momentum $l$ and zero
projection of the angular momentum $m=0$, perturbed by strong
magnetic field ${\cal H}={\cal H}_z$. Let us write expansions of
the energy eigenvalue and the wave function of this state in the
form
 $$ E=\sum^{\infty}_{k=0}E_k\gamma^{2k} \ , \quad \psi(\vec
r)=\sum^{\infty}_{k=0}\psi_k(\vec r)\gamma^{2k} \ , \eqno(1) $$ $$
\gamma=n^3 {\cal H}/{\cal H}_0 \ , \quad {\cal
H}_0=e^3m^2c/\hbar^3=2.35\times 10^9 G \ .
$$
Here $\psi_k$ is the correction of the order $k$ to the Coulomb wave
function, which satisfies the inhomogeneous equation
$$
(\hat{H_0}-E_0)\psi_k=-\hat{H_1}\psi_{k-1}+\sum^k_{j=1}E_j\psi_{k-j} \ ,
\eqno(2)
$$
where
$$
\hat{H_0}=-\frac{1}{2}\nabla^2-\frac{1}{r} \ , \quad
\hat{H_1}=\frac{1}{8}(r^2-z^2) \ .
$$
(We use atomic units.) To change the differential equation by an algebraic
one we introduce the moments of the order $k$
$$
P^k_{\sigma\nu}=\langle\tilde{\psi_0}|r^{\sigma-\nu}z^\nu|\psi_k\rangle \ ,
\eqno(3)
$$
where
$$
|\tilde{\psi_0}\rangle=Ce^{-r/n} , \quad \sigma\ {\rm and}\ \nu\
{\rm are\ integer}\ .
$$
In this definition $|\tilde{\psi_0}\rangle$ contains
only exponential factor of the unperturbed wave function, bearing its
scale and having no nodes. (Note, that the common normalization factor
of all moments can be chosen arbitrary.) Just as it was done in Ref. [2],
multiply Eq. (2) from the left by
$\langle\tilde{\psi_0}|r^{\sigma-\nu}z^\nu$
and use the possibility for the hamiltonian to act to the left, on
explicitly known functions. This way the recurrence relation for
moments of the order $k$ results:
$$ \frac{(\sigma-\nu)(\sigma+\nu+1)}{2}P^k_{\sigma-2,\nu}+
\frac{\nu(\nu-1)}{2}P^k_{\sigma-2,\nu-2}-
\frac{\sigma+1-n}{n}P^k_{\sigma-1,\nu}+P^0_{\sigma\nu}E_k=
R^{k-1}_{\sigma\nu} \ , \eqno(4)
$$
where
$$
R^{k-1}_{\sigma\nu} \equiv
\frac{1}{8}(P^{k-1}_{\sigma+2,\nu}-P^{k-1}_{\sigma+2,\nu+2})-
\sum^{k-1}_{j=1}E_jP^{k-j}_{\sigma\nu} \ .
$$
The right-hand side of Eq. (4) contains moments only of preceding orders.
The $E_k$ coefficient, which we call the hyper-susceptibility of the order
$k$, can also be expressed through the preceding orders moments. This
expression follows from (4) and will be written down below, separately for
each of the cases under investigation.

   Quite similar it is possible to consider another perturbation of
the same state, if this perturbation has the form of a polynomial in
$r$ and $z$. It is enough to change only the right-hand side of Eq. (4)
for this aim. The expression in brackets in the right-hand side,
representing the magnetic field contribution, should be replaced by another
function of the preceding orders moments, created by the new perturbation.

   The succession of the based on relation (4) computations becomes
more lucid if one represents on a plot the lattice of points with
integer coordinates of columns $\sigma$ and rows $\nu$. The indices of
moments, necessary to compute energy and wave function corrections,
are placed on this lattice in the sector $\sigma\ge\nu-1$, \ $\nu\ge0$.
In the general case Eq. (4) relates moments of the order $k$,
indices of which are located in the vertices of a rectangular
triangle (see examples in Fig. 1). When one vertex appears to be on
the line $\sigma=\nu-1$ and another one is outside the above indicated
sector, the triangle (example $A$) transforms into segment (example $B$),
and Eq. (4) turns into relation between two moments from different
rows. Equation (4) relates moments in pairs also along each of the
lines with $\nu=0$ and $\nu=1$ (examples $C$ and $D$). At $k=0$, as the
direct integration indicates, among the moments
$\langle\tilde{\psi_0}|r^{\sigma-\nu}z^\nu|n,l\rangle$
are equal to zero those, for which $l-1\le\sigma<n-2$. If the perturbation
is even, as in the case of  the Zeeman effect, all corrections to the wave
function have the same parity. Therefore, in all PT orders the moments with
odd sum $\nu+l$ vanish.

\bigskip

\begin{center}
\bf
3. ISOLATED STATES
\end{center}

\bigskip

    Magnetic field does not mix states with different parities,
therefore besides the ground state, $2s$-, $2p$- and $3p$-states
should also be considered as non-degenerate. Let us show, how the
moment method works in the last case. Only the moments with odd $\nu$
values can be different from zero. In initial PT order
$$
P^0_{\sigma,2\ae+1}=-\frac{\sigma(\sigma+3)!}{18(2\ae+3)}
\left(\frac{3}{2}\right)^\sigma , \quad \ae\ {\rm is\ integer} \ .
\eqno(5)
$$
For all following orders an additional requirement is introduced:
corrections to the function $|\psi_0\rangle$ should be orthogonal to
the function $|\psi_0\rangle$ itself, i.e.
$\langle\psi_0|\psi_k\rangle=\delta_{0,k}$. This condition
is the routine element of the Brillouin-Wigner perturbation theory (see,
e.g. [20]). It was used in Ref. [2]. In the case of $3p$-state it
takes the form of the following additional relation between the moments
$$
P^k_{11}-\frac{1}{6}P^k_{21}=\delta_{0,k} \ .  \eqno(6)
$$
To obtain an expression for $E_k$, we substitute into recurrence relation
(4) first $\nu=1$, $\sigma=1$, then $\nu=1$, $\sigma=2$. The solution
of the obtained system of two linear equations is
$$
E_k=R^{k-1}_{11}-\frac{1}{6}R^{k-1}_{21}=
\frac{1}{8}(P^{k-1}_{31}-P^{k-1}_{33})-
\frac{1}{48}(P^{k-1}_{41}-P^{k-1}_{43}) \ ,   \eqno(7)
$$
$$
P^k_{01}=5R^{k-1}_{11}-\frac{1}{3}R^{k-1}_{21} \ .  \eqno(8)
$$
The sum, containing hyper-susceptibilities of preceding orders,
dropped out of the final expression for $E_k$ owing to orthogonality
condition (6).

   Equations (4), (6) and (7)  form the closed system of recurrence
relations. In each order $k\ge1$ the sequence of computations is
arranged as follows. First, the coefficient $E_k$ is evaluated with the
help of (7). On the next step $\nu=1$ and $\sigma=3$ are substituted
into recurrence relation (4). In this case together with
orthogonality condition (6) it forms a system of equations from which
initial elements of the row of moments with $\nu=1$ are obtained:
$$
P^k_{11}=\frac{1}{3}(R^{k-1}_{31}-P^0_{31}E_k)=\frac{1}{6}P^k_{21} \ .
$$
Successively increasing $\sigma$ by one, it is not difficult to come
to the necessary boundary moment of this row. Substituting then in
Eq. (4) $\nu=3$ and $\sigma=3$, we get the initial moment of the
next row
$$
P^k_{33}=3(3P^k_{11}-R^{k-1}_{33}+P^0_{33}E_k)
$$
and so on.

   The boundary moments, i.e. the moments with the highest for the
given order $k$ indices $\sigma$ and $\nu$ values are determined by
the following  conditions. To compute the hyper-susceptibility  of
high order $K$ the following moments are required: $P^1_{\sigma\nu}$
from the region $\nu-1\le\sigma\le3K$, \ $0\le\nu\le2K$, then
$P^2_{\sigma\nu}$ from the region $\nu-1\le\sigma\le3K-3$, \
$0\le\nu\le2K-2$ and so on.
   The computation of $E_k$ coefficients for the other isolated
states goes a bit more simple.

\bigskip

\begin{center}
\bf
4. DEGENERATE STATES IN THE MOMENT METHOD
\end{center}

\bigskip

     We consider as an example a pair of splitted by the magnetic field
states $|3s\rangle$ and $|3d\rangle$. Taking into account degeneracy and using
the functions
$$
|\psi_0\rangle=\cos{\alpha}|3s\rangle+\sin{\alpha}|3d\rangle \ , \quad \
|\tilde{\psi_0}\rangle\sim e^{-r/3} \ ,
$$
it is not difficult to obtain the zero order moments
$$
P^0_{\sigma,2\ae}=\frac{(\sigma+2)!}{54(2\ae+1)}
\left(\frac{3}{2}\right)^\sigma
\left(\sigma(\sigma+1)-\frac{\ae(\sigma+3)(\sigma+4)}{2\ae+3}\xi
\right) \ ,   \eqno(9)
$$
where
$$
\xi=\sqrt{2}\:tg\:\alpha \ .
$$
The moments in odd rows are equal to zero in all PT orders,
$P^k_{\sigma,2\ae+1}=0$. The orthogonality condition
$\langle\psi_0|\psi_k\rangle=\delta_{0,k}$ is equivalent to the following
relation between the moments:
$$
12P^k_{10}-\frac{1}{3}(4+\xi)P^k_{20}+\xi P^k_{22}=18P^k_{00} \ , \quad
\ k\ge1 \ . \eqno(10)
$$
There are two independent ways for expressing hyper-susceptibility $E_k$
through preceding orders moments.

     a) Equation (4) at $\nu=0$ and $\sigma=0$ determines the moment
$P^k_{-1,0}$, which is used on the next step in the system of linear
equations. Note that $E_k$ coefficient drops out of Eq. (4) as a
consequence of $P^0_{00}=0$ equality. The system of equations, containing $E_k$
arises if one put in Eq. (4) first $\nu=0$, $\sigma=1$, and then $\nu=0$,
$\sigma=2$. Its solution is
$$
E^{(a)}_k=\frac{9}{2}R^{k-1}_{00}-3R^{k-1}_{10}+\frac{1}{3}R^{k-1}_{20}
\ , \eqno (11a)
$$
$$
P^k_{00}=-9R^{k-1}_{00}+6R^{k-1}_{10}-\frac{1}{3}R^{k-1}_{20} \ .
\eqno(12)
$$
The obtained as a by-product moment $P^k_{00}$ is substituted into the
right-hand side of orthogonality condition (10).

     b) Substituting into Eq. (4) $\nu=2$ and $\sigma=2$ together with
the obtained $P^k_{00}$ value, we get the second independent expression for
$E_k$
$$
E^{(b)}_k=\frac{1}{2}(9R^{k-1}_{00}-6R^{k-1}_{10}+
\frac{1}{3}R^{k-1}_{20}+R^{k-1}_{22}) \ . \eqno(11b)
$$
Equations (11a) and (11b) result in single magnetic susceptibility $E_1$ value
at two $\xi$ values:
$$
\xi=\xi_{1,2}=\frac{-13\pm3\sqrt{41}}{10}  \eqno(13)
$$
It is a natural result, which in Rayleigh-Schr\"odinger PT follows from the
secular equation. In what follows the notation $3s$ is kept for the state with
small admixture of $d$-wave and $\xi=\xi_1=(3\sqrt{41}-13)/10$, and $3d$ denotes
the orthogonal to this state combination of $s$- and $d$-waves. In the
following approximations the unambiguity condition, applied to
hyper-susceptibility of $(k+1)$th order, $E^{(a)}_{k+1}=E^{(b)}_{k+1}$ is
equivalent to relation between seven unknown moments of the order $k$:
$$
\xi[9(P^k_{20}-P^k_{22})-6(P^k_{30}-P^k_{32})]-
\frac{1}{3}(1-2\xi)P^k_{40}+
\frac{2}{3}(1+\xi)P^k_{42}-P^k_{44}=
$$
$$
8[\xi(9S^k_{00}-6S^k_{10})-\frac{1}{3}(1-2\xi)S^k_{20}+S^k_{22}] \ .
\eqno(14)
$$
Here $S^k_{\sigma\nu}=\sum\limits^{k}_{j=1}E_jP^{k-j}_{\sigma\nu}$ and
$\xi=\xi_{1,2}$. One more constraint on the moments is orthogonality condition
(10). To obtain a closed system, Eqs. (10) and (14) should be supplemented
by seven equations, following from recurrence relation (4). The set of
unknown variables includes $P^k_{10}, P^k_{20}, P^k_{30}, P^k_{40},
P^k_{22}, P^k_{32}, P^k_{42}, P^k_{34}, P^k_{44}$. It is enough to determine
only two moments $P^k_{10}$ and $P^k_{22}$ from the system of nine equations.
Then, with the help of already known moments and relation (4) it is not
difficult to compute successively all necessary moments of the given order,
passing line by line the lattice of indices, like in the case of $3p$-state.

     Quite similar, it is possible to accomplish the computation for a state
with arbitrary $n$ value and zero projection of angular momentum. The
unperturbed wave function has definite parity and contains $g$ degenerate
in energy terms. Therefore $g-1$ independent mixing parameters explicitly enter
the zero order moments. There are two groups of moments in every order $k\ge1$.
Recurrence relation (4) connects the moments $P^k_{\sigma\nu}$ with
$\sigma<n-2$ and separately the moments of the same order but with
$\sigma\ge n-2$. Moments from different groups are connected by recurrence
relation only through the moments of preceding orders.

     To obtain all independent expressions for $E_k$, one should substitute
into recurrence relation (4) the successively increasing $\nu$ values of the
given parity and corresponding set of $\sigma$ values:
$$
0\le\nu\le\sigma\le n-1 \ .
$$
At every $\nu$ value the unambiguously solvable system of equations is obtained.
Its solution contains independent expression for $E_k$ and a set of moments of
the order $k$, to be substituted in analogous system at the next $\nu$ value.
Thus, $g$ independent expressions for $E_k$ in terms of the preceding orders
moments result. The unambiguity condition of $E_1$ determines $g-1$ mixing
coefficients. The unambiguity condition of $E_{k+1}$ at $k\ge1$ allows to
express the moments $P^k_{\sigma\nu}$ from the domain $\sigma\ge n-2$ through
preceding orders moments. The equations expressing the unambiguity of $E_{k+1}$
are supplemented by orthogonality condition
$\langle\psi_0|\psi_k\rangle=0$ and by necessary number of equations, obtained
from recurrence relation (4), to close the system.

\bigskip

\begin{center}
\bf
5. RESULTS
\end{center}

\bigskip

{\it Energy eigenvalues}
\bigskip

     For all levels with $n\le3$ and $m=0$ with the help of the moment method
we have obtained Zeeman's hyper-susceptibilities $E_k$ up to 75th order,
see Table 1. All computations were carried out with 32 decimal digits. Complete
agreement is observed with the results of Ref. [8] and Ref. [7],
containing first five coefficients $E_k$ for the ground state and three initial
coefficients for both $2s$- and $2p$-states in the form of rational fractions.
In Ref. [9] a difference was detected between the obtained in this work
expression for $E_2$ coefficient and its value at $l=1$ in Ref. [7]. This
deviation is confirmed. As it follows from expression of Ref. [9],
$E^{(2p)}_2=-45.556$, but our result is $E^{(2p)}_2=-42$ in agreement with [7].

     Energy eigenvalues $E(\gamma)$ of six states, obtained by corresponding
power series (PS) summation with the help of Pad\'e approximants
$[L/L](\gamma^2)$ and $[L/L-1](\gamma^2)$ are shown in Figs. 2a-2c. These
figures represent also the region of convergence of Pad\'e approximants.
Without expansion in ${\cal H}^2$, by means of the splines method, which is one
of modifications of the variational method, energy eigenvalues of states under
consideration were computed for some ${\cal H}$ values in Ref. [21]. They are
also indicated in Figs. 2a-2c. Reference [21] has the best precision among all
non-perturbative calculations and contains a comparison of a large number of
previous computations. The precision of PS summation with the help of Pad\'e
approximants is high enough. At $\gamma\approx1$ for $2s$- and $3p$-states
three or four stable digits of energy eigenvalue are established, and at least
two decimal digits for energies of the other states are obtained. Results of
PT series summation together with some results of Ref. [21] are represented
in Table 2. This table also illustrates convergence of Pad\'e approximants
we used. The convergence is sharply increased with $\gamma$ decreasing, and
at $\gamma<0.3$ the precision of PS sum exceeds the precision of variational
calculations [21].

     The PT coefficients $E_k$ approach, as the order increases, to the
asymptotic, the leading term of which for the Zeeman effect is [4,5]:
$$
\tilde{E_k}=(-1)^{k+1}\frac{D_{nl}}{\pi^{2n+1/2}}
\left(\frac{n^2}{\pi}\right)^{2k}\Gamma(2k+2n-1+\frac{(-1)^l}{2}) \ .
\eqno(15)
$$
For the levels under consideration
$$
D_{1s}=32 \ , \ D_{2s}=128 \ , \ D_{2p}=64 \ ,
$$
$$
\ D_{3s}=\frac{2^{15}}{3^4}\left(\alpha_1-\frac{\alpha_2}{2\sqrt{2}}\right)^2
\ ,
\ D_{3d}=\frac{2^{15}}{3^4}\left(\alpha_2+\frac{\alpha_1}{2\sqrt{2}}\right)^2
\ ,
$$
where
$$
\alpha_1=-\left(\frac{1}{2}+\frac{13}{6\sqrt{41}}\right)^{1/2} \ ,
\ \alpha_2=\left(\frac{1}{2}-\frac{13}{6\sqrt{41}}\right)^{1/2} \ .
$$
This result was obtained by the method introduced by Bender and Wu [22].
First, the penetreability of the barrier at imaginary magnetic field value was
computed in quasiclassical approximation, then the dispersion relation in
${\cal H}^2$ was applied. Stricktly speaking, according to the conditions of
Ref. [5], expression (15) is not referred to the case of $3p$-state. But it
is reasonable to suppose that formula (15) describes all six discussed here
states with coefficient $D_{3p}$ fitted by comparing $E_k$ and $\tilde{E_k}$ of
sufficiently high orders. The result is $D_{3p}=2^{13}/3^3$. The approach of
exact coefficients $E_k$ to asymptotic (15) is illustrated by Fig. 3.

     In Refs. [4,5] the corrections to the asymptotic $\tilde{E_k}$ were
obtained for some states, among which $2s$-, $2p$- and $3p$-states were
absent. Writing the corrections as
$$
\frac{E_k}{\tilde{E_k}}=c_0+\frac{c_1}{2k}+\frac{c_2}{(2k)^2}+... \ ,
\eqno(16)
$$
it is not difficult, following the method of Ref. [22], to obtain $c_i$
coefficients for all missing in [4,5] states, see Table 3. As the number of
$c_i$ coefficients included in Eq. (16) is increased, the precision of their
determination increases as well.

     For $3p$-state this stability of the power correction coefficients
confirms that the leading term $\tilde{E_k}$ of the asymptotic is determined
correctly. Note, that due to dispersion relation in ${\cal H}^2$ [5]
coefficients $c_i$ are related with the corrections to quasiclassical
approximation for the barrier penetreability at ${\cal H}^2<0$. The
straightforward computation of quasiclassical corrections is a complicated
enough problem.

\bigskip

{\it Wave functions}
\bigskip

     By analogy with the anharmonic oscillator [2] case, the correction
$|\psi_k\rangle$ to the Coulomb wave function has the form of a polynomial
in $r$ and $\cos\theta$, multiplied by $|\tilde{\psi_0}\rangle$. The
perturbation is a polynomial and the operator in the left-hand side of
Eq. (2) does not change the suggested structure of $|\psi_k\rangle$. For
the ground state
$$
|\psi_k\rangle = (\sum^k_{j=0}\sum^{3k}_{i=2j}a^{(k)}_{ij}r^i
\cos^{2j}\theta)|\tilde{\psi_0}\rangle\equiv
(\sum^k_{j=0}\sum^{3k}_{i=2j}a^{(k)}_{ij}r^{i-2j}z^{2j})
|\tilde{\psi_0}\rangle \ . \eqno(17)
$$
A remark about the origin of the summation boundaries should be done. The
$\theta$ dependence is introduced only by the expression $r^2\cos^2\theta$,
contained in $\hat{H_1}$, therefore minimal power of radius in the internal
sum of expression (17) coincides with the power of $\cos\theta$, and the
highest power of $\cos^2{\theta}$ coincides with PT order. It is possible to
check with the help of Eq. (2) that the highest power of radius in
$|\psi_k\rangle$ is bigger by three than that in $|\psi_{k-1}\rangle$. A system
of linear equations determining coefficients $a^{(k)}_{ij}$ follows from (17)
and looks like
$$
\sum^k_{j=0}\sum^{3k}_{i=2j}P^0_{i+\alpha,j+\beta}a^{(k)}_{ij}=
P^k_{\alpha\beta} \ , \eqno(18)
$$
$$
2\beta\le\alpha\le3k \ , \ 0\le\beta\le k \ .
$$
In place of the indicated $\alpha$ and $\beta$ values one can choose another
their set, resulting in $(2k+1)(k+1)$ independent equations. This possibility
is useful to check the precision of computations. We checked also the
orthogonality of the obtained corrections (17) to the function
$|\psi_0\rangle$. In our computation the orthogonality condition was preserved
with a reasonable precision up to 18th PT order. Corrections to the wave
functions up to second order are given in Ref. [7]. By comparing this work
and our one the single point of deviation was observed. The sign at the
$r^3\gamma^2$ term in Ref. [7] is erroneous and as a result the first order
correction is not orthogonal to unperturbed wave function in work [7].

     Within the moment method we have computed $|\psi(0)|^2$ values by summing
with the help of Pad\'e approximants the PT series for the normalization factor
and for the wave function itself. Results are represented in Fig. 4. One can
see that 18 PT orders allow to advance up to $\gamma\approx0.4$. Besides,
there is an agreement with Ref. [23], where the wave function of the ground
state was computed by a different method.

\bigskip

\begin{center}
\bf
6. CONCLUSIONS
\end{center}

\bigskip

     A good agreement between energy eigenvalues obtained by PT series
summation and the corresponding results of independent variational calculations
provides a twofold verification. First, it is confirmed that the computed PT
coefficients are correct. Then, the absence of non-perturbative contributions
to energy eigenvalues is indicated (the summation method is chosen correctly).

     Our investigation demonstrates the high efficiency of the
Ader moment method. Let us stress once more the properties of the method which
allowed to obtain the high PT orders for the Zeeman effect. These properties
could be useful for solving some other problems of the perturbation theory.

     -- The simple enough recurrence relations are obtained without variables
separation in the initial equation. With their aid the corrections computation
is reduced to purely algebraic procedure.

     -- The nodes of the wave function are not shown explicitly and do not
complicate the computations. As a consequence the excited states can be
considered.

     -- The level degeneracy, as was demonstrated above, does not restrict
application of the moment method, but it leads to more complicated sequence
of operations when solving the recurrence relations.

     -- Besides, the set of moments used for hyper-susceptibilities
determination contains information about the wave function. Corrections to
Coulomb wave function, just as corrections to energy eigenvalues, follow from
purely algebraic procedure.

     As it was already mentioned, application of the moment method to the
hydrogen atom is not restricted by the Zeeman effect case. Any perturbation of
the kind $V(r)=\sum b_{ij}r^{i-j}z^j$ leads to recurrence relation (4) with
an obvious simple modification of the right-hand side. For example,
this way it is possible to take into account the joint influence of external
electric and magnetic fields, homogeneous or consisting of finite number of
multipoles.

\bigskip

\begin{center}
\bf
ACKNOWLEDGMENTS
\end{center}

\bigskip

     We would like to express our deep gratitude to V.S.Popov for critical
reading the manuscript and for some valuable comments, leading to its
improvement. We are also grateful to V.G.Ksenzov for useful discussion and to
V.D.Mur and B.M.Karnakov for their interest to our work. One of the authors
(V.A.G.) would like to thank V.L.Morgunov and A.A.Panfilov for helpful
discussions of questions related to numerical calculations.

\newpage

\begin{center}
{\bf Table 1.}
Hyper-susceptibilities of hydrogen atom in magnetic field.
\end{center}
\vspace{2mm}ÿ 
\begin{center}
\begin{tabular}{|c|l|l|}
\hline
\multicolumn{2}{|c|}{$1s-state$} & \multicolumn{1}{|c|}{$2s-state$} \\
\hline
\multicolumn{1}{|c|}{$k$} & \multicolumn{1}{|c|}{ $E_k$ } &
\multicolumn{1}{|c|}{ $E_k$ } \\
\hline
 1 & $+2.50000000000000000000000000000\times10^{-1}$ &
 $+3.50000000000000000000000000000\times10^{0}$ \\
 2 & $-2.76041666666666666666666666667\times10^{-1}$ &
 $-1.59333333333333333333333333333\times10^{2}$ \\
 3 & $+1.21115451388888888888888888889\times10^{0}$ &
 $+2.25084444444444444444444444444\times10^{4}$ \\
 4 & $-9.75540590639467592592592592593\times10^{0}$ &
 $-5.51663442962962962962962962963\times10^{6}$ \\
 5 & $+1.17863024612238377700617283951\times10^{2}$ &
 $+1.88165092810271604938271604938\times10^{9}$ \\
 6 & $-1.95927276058352435076678240741\times10^{3}$ &
 $-8.20439178651205530864197530864\times10^{11}$ \\
 7 & $+4.27486169952196866486628589997\times10^{4}$ &
 $+4.38959673572860094025228856975\times10^{14}$ \\
 8 & $-1.18693528256085740621952195187\times10^{6}$ &
 $-2.81689767929056062290444312309\times10^{17}$ \\
 9 & $+4.09726018688028425780092822833\times10^{7}$ &
 $+2.13733235781748489849361215413\times10^{20}$ \\
10 & $-1.72515623494757933487367089115\times10^{9}$ &
 $-1.89790071689337590593564645501\times10^{23}$ \\
11 & $+8.71666539327097376193289896759\times10^{10}$ &
 $+1.95582352962753949948044058249\times10^{26}$ \\
12 & $-5.21094093401193811088710286758\times10^{12}$ &
 $-2.32142997906702906114575376593\times10^{29}$ \\
13 & $+3.64053240123290947096878491058\times10^{14}$ &
 $+3.15101443471988021776783108519\times10^{32}$ \\
14 & $-2.94037039347649323969534446706\times10^{16}$ &
 $-4.85797163582017764548476077630\times10^{35}$ \\
15 & $+2.71957243076911667415082196285\times10^{18}$ &
 $+8.45204454588950406643659448685\times10^{38}$ \\
20 & $-1.08008064080683361642422438535\times10^{29}$ &
 $-6.77777320558482045836185652856\times10^{55}$ \\
30 & $-1.60767231445546160409172754060\times10^{53}$ &
 $-2.69333119877780660677693447444\times10^{92}$ \\
40 & $-1.86009993885524622493595668510\times10^{80}$ &
 $-6.32192703798840441487069765398\times10^{131}$ \\
50 & $-3.13170157718318015959704309179\times10^{109}$ &
 $-1.86940626135923124849582995810\times10^{173}$ \\
60 & $-2.82752607152079516815460149212\times10^{140}$ &
 $-2.71176213794760523959296668884\times10^{216}$ \\
70 & $-7.04995099684986774958800182910\times10^{172}$ &
 $-1.02247285421591824260835479011\times10^{261}$ \\
75 & $+3.31500360451477843327480892738\times10^{189}$ &
 $+5.81143908209506920438376327293\times10^{283}$ \\
\hline
\end{tabular}
\end{center}

\newpage

\begin{center}
{\bf Table 1.}
Continuation A.
\end{center}
\vspace{2mm}
\begin{center}
\begin{tabular}{|c|l|l|}
\hline
\multicolumn{2}{|c|}{$2p-state$} & \multicolumn{1}{|c|}{$3s-state$} \\
\hline
\multicolumn{1}{|c|}{$k$} & \multicolumn{1}{|c|}{ $E_k$ } &
\multicolumn{1}{|c|}{ $E_k$ } \\
\hline
 1 & $+1.50000000000000000000000000000\times10^{0}$ &
 $+1.95785147671119547722992448839\times10^{1}$ \\
 2 & $-4.20000000000000000000000000000\times10^{1}$ &
 $-7.99255848864256699334910438169\times10^{3}$ \\
 3 & $+4.24000000000000000000000000000\times10^{3}$ &
 $+9.86542519182491373199618234956\times10^{6}$ \\
 4 & $-7.43649066666666666666666666667\times10^{5}$ &
 $-2.09268850909736967372029449066\times10^{10}$ \\
 5 & $+1.87097828977777777777777777778\times10^{8}$ &
 $+5.88259296456916749962643192656\times10^{13}$ \\
 6 & $-6.23210518955750264550264550265\times10^{10}$ &
 $-2.03491551010345668150921550764\times10^{17}$ \\
 7 & $+2.63615589562176770773494583018\times10^{13}$ &
 $+8.31244877905745777735770437889\times10^{20}$ \\
 8 & $-1.38041101245356842308845936964\times10^{16}$ &
 $-3.91936927735269422610169429796\times10^{24}$ \\
 9 & $+8.78777021937730413143094721050\times10^{18}$ &
 $+2.10554214073622864491511736652\times10^{28}$ \\
10 & $-6.70289485780879092143754218449\times10^{21}$ &
 $-1.27856250168620586782771927856\times10^{32}$ \\
11 & $+6.04903613875484403654574962209\times10^{24}$ &
 $+8.73019078199070904206045207631\times10^{35}$ \\
12 & $-6.38629370869963249560760013166\times10^{27}$ &
 $-6.67712550914959188602686231758\times10^{39}$ \\
13 & $+7.80769206270237413951487473753\times10^{30}$ &
 $+5.70134624264532995993336408500\times10^{43}$ \\
14 & $-1.09524180989414207944239838172\times10^{34}$ &
 $-5.41744227892988968803716808669\times10^{47}$ \\
15 & $+1.74827746288164542821561289029\times10^{37}$ &
 $+5.70970876646976596904731808022\times10^{51}$ \\
20 & $-9.93004508733968922803863745812\times10^{53}$ &
 $-3.11636589919914001005211784216\times10^{72}$ \\
30 & $-2.49139318960793926600959408638\times10^{90}$ &
 $-3.53759135634213416803068610350\times10^{116}$ \\
40 & $-4.27118446981052432109181315161\times10^{129}$ &
 $-1.74975503131501618694229226731\times10^{163}$ \\
50 & $-9.94612680182593582345101783873\times10^{170}$ &
 $-9.32290913905465430025588878862\times10^{211}$ \\
60 & $-1.18983403811968620569412506832\times10^{214}$ &
 $-2.21489826289171591863053880143\times10^{262}$ \\
70 & $-3.81690935673691049129436112647\times10^{258}$ &
 $-1.28257798832278036045229086676\times10^{314}$ \\
75 & $+2.01879251467438954569448392302\times10^{281}$ &
 $+2.80538297459908381755301158907\times10^{340}$ \\
\hline
\end{tabular}
\end{center}

\newpage

\begin{center}
{\bf Table 1.}
Continuation B.
\end{center}
\vspace{2mm}
\begin{center}
\begin{tabular}{|c|l|l|}
\hline
\multicolumn{2}{|c|}{$3p-state$} &
\multicolumn{1}{|c|}{$3d-state$} \\
\hline
\multicolumn{1}{|c|}{$k$} & \multicolumn{1}{|c|}{ $E_k$ } &
\multicolumn{1}{|c|}{ $E_k$ } \\
\hline
 1 & $+9.00000000000000000000000000000\times10^{0}$ &
 $+5.17148523288804522770075511605\times10^{0}$ \\
 2 & $-3.53109375000000000000000000000\times10^{3}$ &
 $-1.01742588635743300665089561831\times10^{3}$ \\
 3 & $+3.44813867578125000000000000000\times10^{6}$ &
 $+6.44144210626019957028207894340\times10^{5}$ \\
 4 & $-5.44958117499880371093750000000\times10^{9}$ &
 $-8.64776860916338232383320244317\times10^{8}$ \\
 5 & $+1.17571125958104512054443359375\times10^{13}$ &
 $+1.64145868005800755132947440105\times10^{12}$ \\
 6 & $-3.18064888985252133830423028128\times10^{16}$ &
 $-4.16016007587619392285546026342\times10^{15}$ \\
 7 & $+1.03477309951815151989739404812\times10^{20}$ &
 $+1.35506144668665011578479155203\times10^{19}$ \\
 8 & $-3.95890060949951787171702519397\times10^{23}$ &
 $-5.45740994533040798137376086700\times10^{22}$ \\
 9 & $+1.75802363547448200866475465911\times10^{27}$ &
 $+2.63004738948388301397471581365\times10^{26}$ \\
10 & $-8.98508915233072068697058004551\times10^{30}$ &
 $-1.47853891447648836002155944574\times10^{30}$ \\
11 & $+5.25232959291187236720839982517\times10^{34}$ &
 $+9.52631535621939194061435526295\times10^{33}$ \\
12 & $-3.49297501706658104776470560233\times10^{38}$ &
 $-6.95353066637932651872818015613\times10^{37}$ \\
13 & $+2.62940716668212328895716830967\times10^{42}$ &
 $+5.70612739572397688871373383049\times10^{41}$ \\
14 & $-2.22927146751916684798616723449\times10^{46}$ &
 $-5.23539432305661143784168964358\times10^{45}$ \\
15 & $+2.11808048348488483409151866222\times10^{50}$ &
 $+5.34692412323682373417129879084\times10^{49}$ \\
20 & $-7.67756482076312202465482219049\times10^{70}$ &
 $-2.59058872169806959065242368765\times10^{70}$ \\
30 & $-5.13502619416402678854105232212\times10^{114}$ &
 $-2.58371178967762066018120986200\times10^{114}$ \\
40 & $-1.78804617789836253655584492158\times10^{161}$ &
 $-1.19414817113404873895159768413\times10^{161}$ \\
50 & $-7.33295913933863525514078947399\times10^{209}$ &
 $-6.10209745061797542849682305696\times10^{209}$ \\
60 & $-1.41449923722599152753674938951\times10^{260}$ &
 $-1.40914942870226252773832277963\times10^{260}$ \\
70 & $-6.89053783630996102530832165310\times10^{311}$ &
 $-7.99404210639204636195381081171\times10^{311}$ \\
75 & $+1.39614310998209220877154524815\times10^{338}$ &
 $+1.73411445429769522609855678396\times10^{338}$ \\
\hline
\end{tabular}
\end{center}

\newpage

{\bf Table 2.}
Convergence of Pad\'e approximants for energy levels of hydrogen atom in magnetic
field. The values taken for comparison from Ref. [21] are marked by $^{*)}$.
\vspace{3mm}
\begin{center}
\begin{tabular}{|c|c|c|c|c|c|}
\hline
\multicolumn{2}{|c|}{$\gamma$} & \multicolumn{2}{|c|}{$0.1$} &
\multicolumn{2}{|c|}{$0.4$}\\
\hline
 & $L$ & $[L/L](\gamma^2)$ & $[L/L-1](\gamma^2)$ &
$[L/L](\gamma^2)$ & $[L/L-1](\gamma^2)$ \\
\hline
             & 21 & $-0.497526480401260$ & $-0.497526480401090$
             & $-0.464665$ & $-0.464599$ \\
             & 22 & $-0.497526480401200$ & $-0.497526480401092$
             & $-0.464658$ & $-0.464600$ \\
$|1s\rangle$ & 23 & $-0.497526480401163$ & $-0.497526480401093$
             & $-0.464653$ & $-0.464601$ \\
             & 24 & $-0.497526480401140$ & $-0.497526480401093$
             & $-0.464648$ & $-0.464601$ \\
             & 25 & $-0.497526480401125$ & $-0.497526480401094$
             & $-0.464644$ & $-0.464602$ \\
\multicolumn{2}{|c|}{ } & \multicolumn{2}{|c|}{ } & \multicolumn{2}{|c|}{ } \\
\hline
\multicolumn{2}{|c|}{$\gamma$} & \multicolumn{2}{|c|}{$0.16$} &
\multicolumn{2}{|c|}{$0.8$} \\
\hline
             & 21 & $-0.1236241775347995$ & $-0.1236241775347930$
             & $-0.098160$ & $-0.098074$ \\
             & 22 & $-0.1236241775347966$ & $-0.1236241775347925$
             & $-0.098153$ & $-0.098074$ \\
$|2s\rangle$ & 23 & $-0.1236241775347951$ & $-0.1236241775347928$
             & $-0.098147$ & $-0.098077$ \\
             & 24 & $-0.1236241775347942$ & $-0.1236241775347928$
             & $-0.098142$ & $-0.098078$ \\
             & 25 & $-0.1236241775347937$ & $-0.1236241775347928$
             & $-0.098137$ & $-0.098079$ \\
\multicolumn{2}{|c|}{ } & \multicolumn{2}{|c|}{$-0.12362418^{\ *)}$} &
\multicolumn{2}{|c|}{$-0.0980892^{\ *)}$} \\
\hline
\multicolumn{2}{|c|}{$\gamma$} & \multicolumn{2}{|c|}{$0.27$} &
\multicolumn{2}{|c|}{$1.08$} \\
\hline
             & 21 & $-0.05468786997811$ & $-0.05468786997805$
             & $-0.045436$ & $-0.045412$ \\
             & 22 & $-0.05468786997796$ & $-0.05468786997784$
             & $-0.045430$ & $-0.045398$ \\
$|3p\rangle$ & 23 & $-0.05468786997795$ & $-0.05468786997790$
             & $-0.045428$ & $-0.045404$ \\
             & 24 & $-0.05468786997793$ & $-0.05468786997788$
             & $-0.045425$ & $-0.045402$ \\
             & 25 & $-0.05468786997791$ & $-0.05468786997789$
             & $-0.045423$ & $-0.045404$ \\
\multicolumn{2}{|c|}{ } & \multicolumn{2}{|c|}{$-0.05468787^{\ *)}$} &
\multicolumn{2}{|c|}{$-0.04540638^{\ *)}$} \\
\hline
\end{tabular}
\end{center}
\vspace{10mm}
\begin{center}
{\bf Table 3.}
Coefficients $c_i$ of the power corrections to asymptotics $\tilde{E_k}$ of
hydrogen atom hyper-susceptibilities.\\
\end{center}
\begin{center}
\begin{tabular}{|c|l|l|l|l|}
\hline
$state$ & \multicolumn{1}{|c|}{$c_1$} & \multicolumn{1}{|c|}{$c_2$} &
\multicolumn{1}{|c|}{$c_3$} & \multicolumn{1}{|c|}{$c_4$}\\
\hline
$1s$ & $-2.61829$ & $+1.282$ & $-2.6$ & $-11$ \\
$2s$ & $-8.938$ & $+37.44$ & $-121$ & $+2.7\times10^2$ \\
$2p$ & $-4.6065$ & $+8.24$ & $-14.3$ & $-4$ \\
$3p$ & $-11.227$ & $+59.5$ & $-239$ & $+6\times10^2$ \\
\hline
\end{tabular}
\end{center}

\newpage

\begin{center}
\bf
FIGURE CAPTIONS
\end{center}

\bigskip

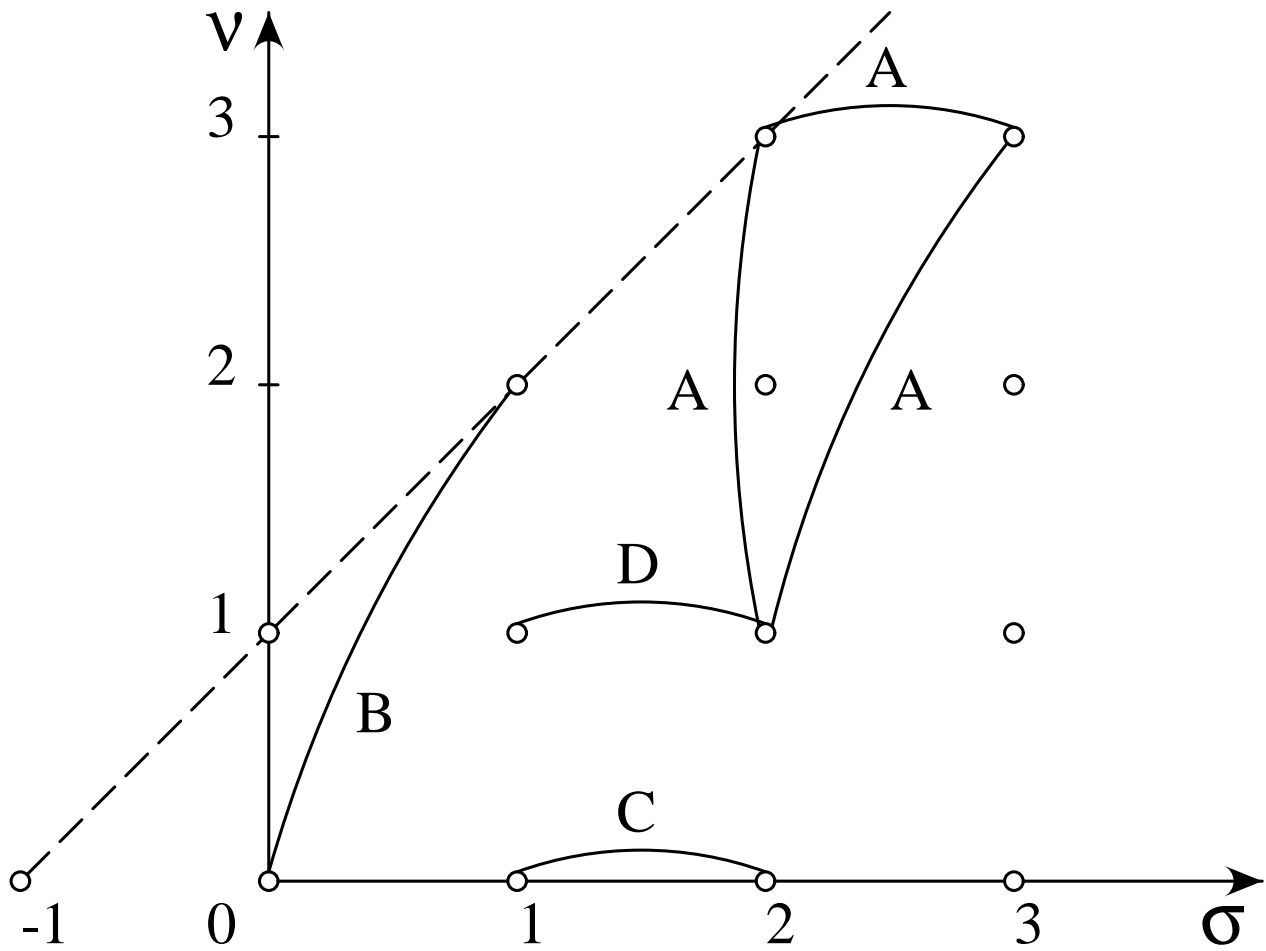
{\bf Fig. 1.}
Indices of moments of order $k$ that are related by means of the main
recurrence relation. Each of the links $A$-$D$ represents one of particular
cases.

\bigskip

{\bf Fig. 2.}
Summation of the PT series for energy levels with the help of Pad\'e
approximants.
$E(\gamma)=[25/25](\gamma^2)$ -- solid curves,
$E(\gamma)=[25/24](\gamma^2)$ -- dashed curves.
Crosses represent results from Ref. [21].

\bigskip

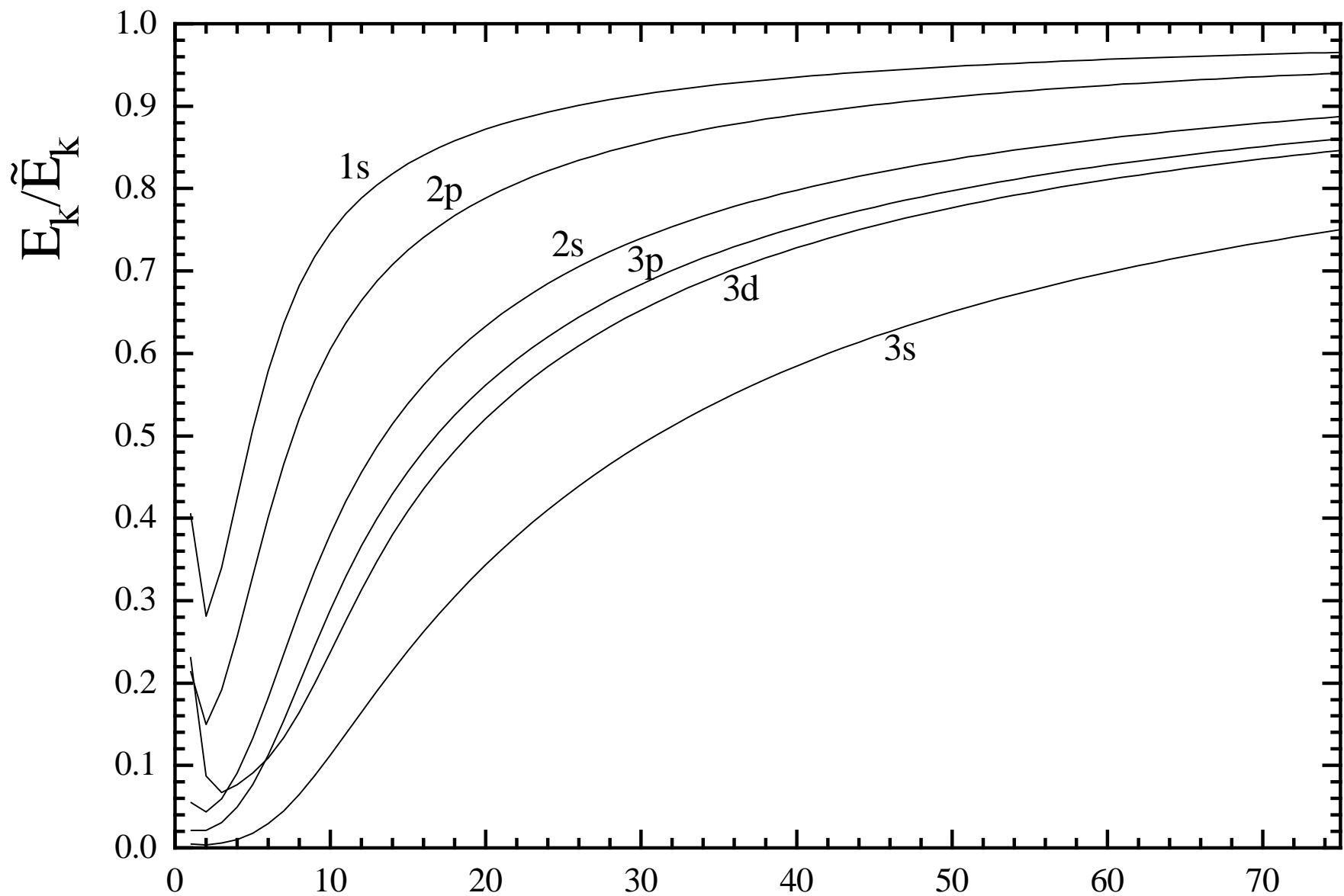
{\bf Fig. 3.}
Approach of exact hyper-susceptibilities $E_k$ to their asymptotics
$\tilde{E_k}$ for the six states of hydrogen atom.

\bigskip

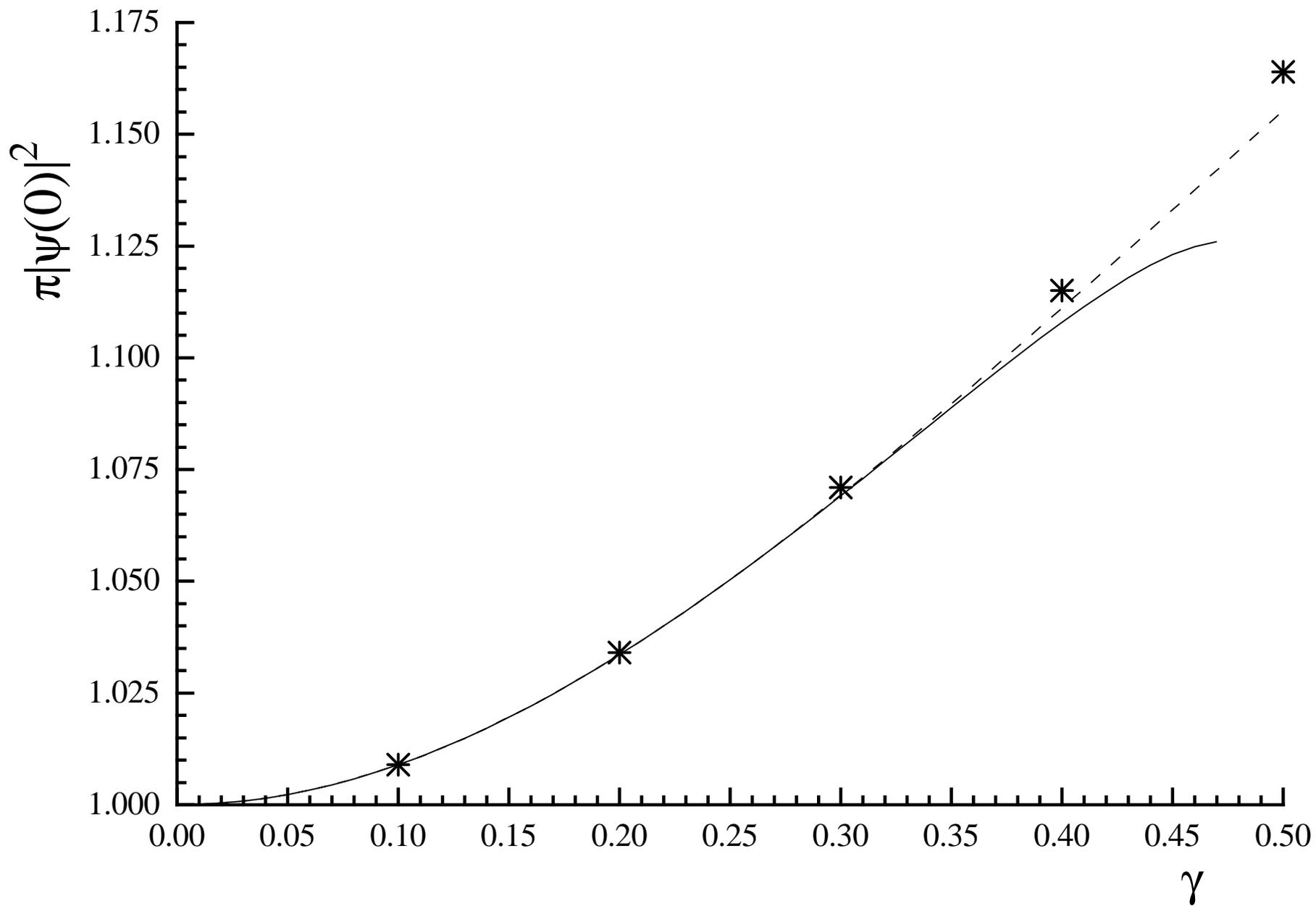
{\bf Fig. 4.}
Calculation of the ground state wave function at the origin $r=0$ by
the moment method. Solid curve was obtained with the help of Pad\'e
approximant $[9/9](\gamma^2)$, dashed curve -- with the help of
$[9/8](\gamma^2)$. Stars denote results of Ref. [23].

\newpage

\begin{center}
\bf
REFERENCES
\end{center}

\bigskip

\begin{itemize}
\item[1.] V.S.Popov, A.V.Sergeev, Pis'ma Zh. Eksp. Teor. Fiz., {\bf 63},
          398 (1996).
\item[2.] J.P.Ader, Phys. Lett. {\bf 97A}, 178 (1983). 
\item[3.] D.Z.Goodson, D.R.Herschbach, Phys. Rev. Lett., {\bf 58}, 1628 (1987).
\item[4.] B.G.Adams et al., Phys. Rev., {\bf 21}, 1914 (1980).
\item[5.] J.E.Avron, Ann. of Phys., {\bf 131}, 73 (1981).
\item[6.] J.\v Ci\v zek, E.R.Vrscay, Int. J. Quant. Chem., {\bf 21}, 27 (1982).
\item[7.] A.Galindo, P.Pascual, Nuovo Cimento, {\bf 34B}, 155 (1976). 
\item[8.] B.R.Johnson, K.F.Scheibner, D.Farrelly,
   Phys. Rev. Lett., {\bf 51}, 2280 (1983). 
\item[9.] A.V.Turbiner, Zs. Phys., {\bf A308}, 111 (1982). 
\item[10.] A.V.Turbiner, Zh. Eksp. Teor. Fiz., {\bf 84}, 1329 (1983). 
\item[11.] V.S.Polikanov, Zh. Eksp. Teor. Fiz., {\bf 52}, 1326 (1967). 
\item[12.] V.S.Pekar, Teor. Mat. Fiz., {\bf 9}, 140 (1971). 
\item[13.] A.D.Dolgov, V.S.Popov, Phys. Lett., {\bf B86}, 185 (1979).
\item[14.] Y.Aharonov, C.K.Au, Phys. Rev., {\bf A20}, 2245 (1979); {\bf A22},
    328 (1980). 
\item[15.] V.Privman, Phys. Rev., {\bf A22}, 1833 (1980). 
\item[16.] V.L.Eletsky, V.S.Popov, Dok. Akad. Nauk SSSR,
           {\bf 250}, 74 (1980);\\
     S.P.Alliluev, V.L.Eletsky, V.S.Popov, Phys. Lett., {\bf A73}, 103 (1979).
\item[17.] S.P.Alliluev, V.M.Weinberg, V.L.Eletsky, V.S.Popov,
    Zh. Eksp. Teor. Fiz., {\bf 82}, 77 (1982). 
\item[18.] R.J.Svenson, S.H.Danforth, Journ. of Chem. Phys., {\bf 57},
    1734 (1972). 
\item[19.] J.Killingbeck, Phys. Lett., {\bf 65A}, 87 (1978). 
\item[20.] J.M.Ziman, {\it Elements of advanced quantum theory}, Cambridge
           (1969).
\item[21.] Jang-Huar Wang, Chen-Shiung Hsue, Phys. Rev., {\bf A52},
    4508 (1995). 
\item[22.] C.M.Bender, T.T.Wu, Phys. Rev., {\bf D7}, 1620 (1973). 
\item[23.] D.Cabib, E.Fabri, G.Fiorio, Nuovo Cimento, {\bf 10}, 185 (1972).
\end{itemize}
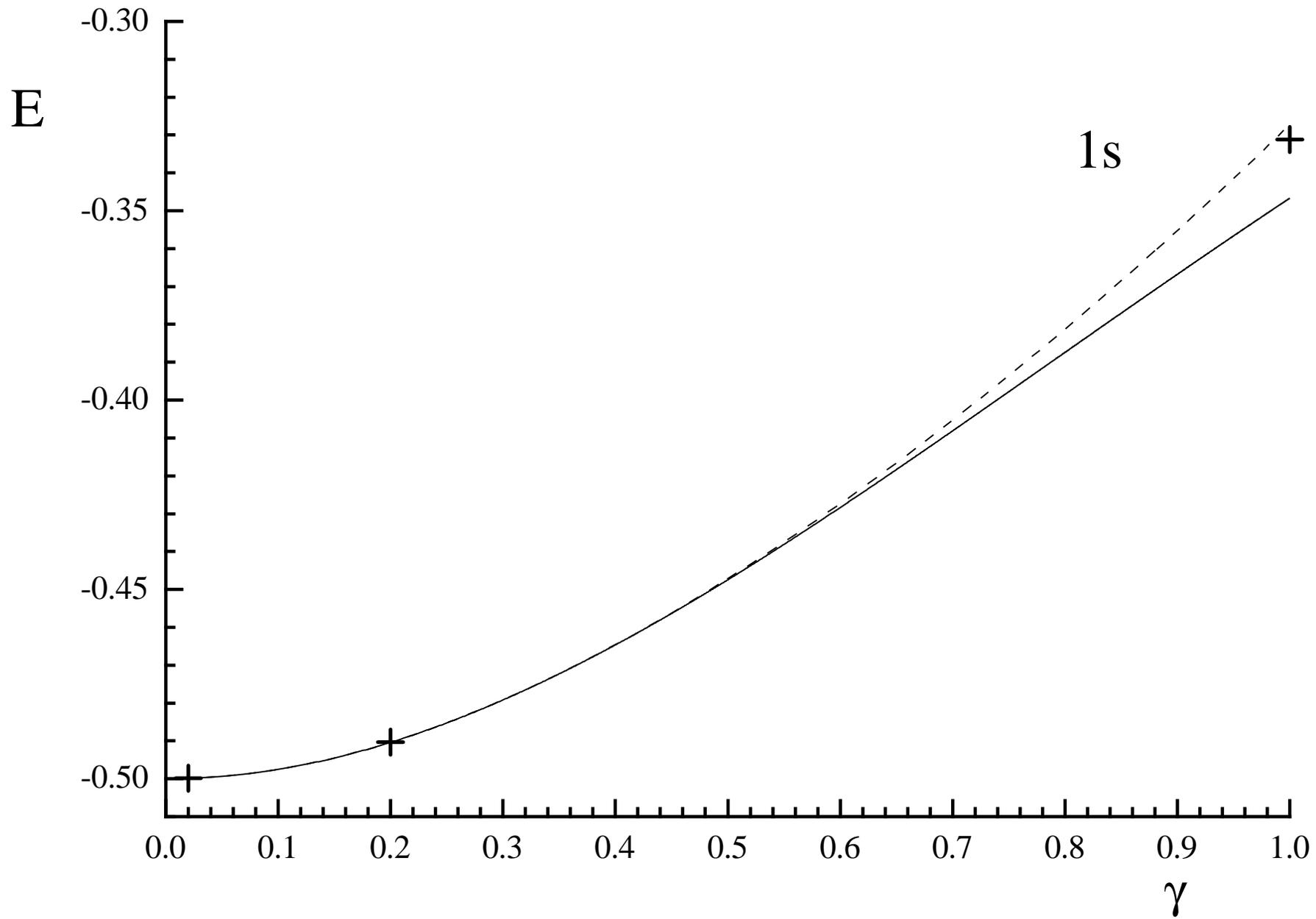
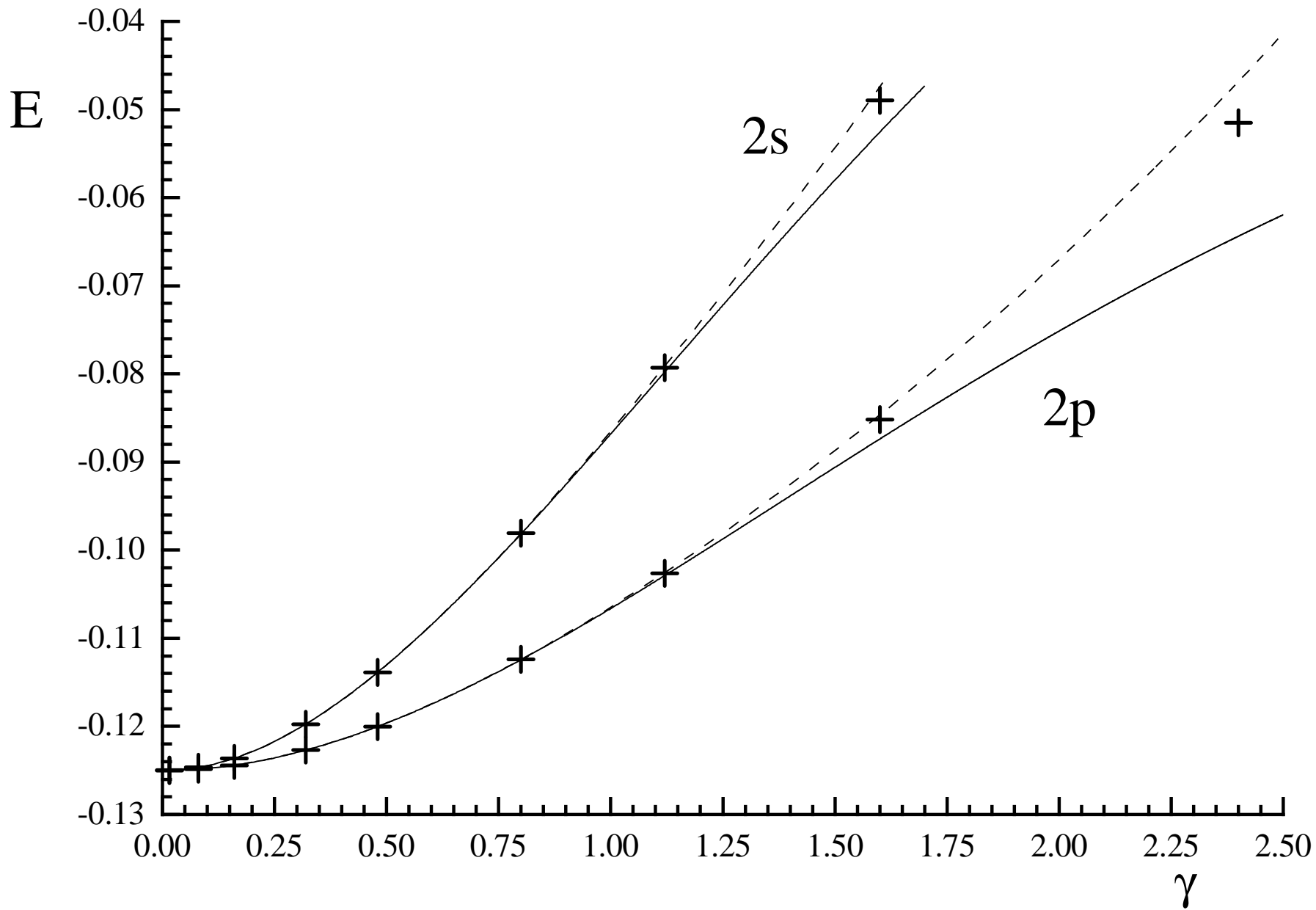
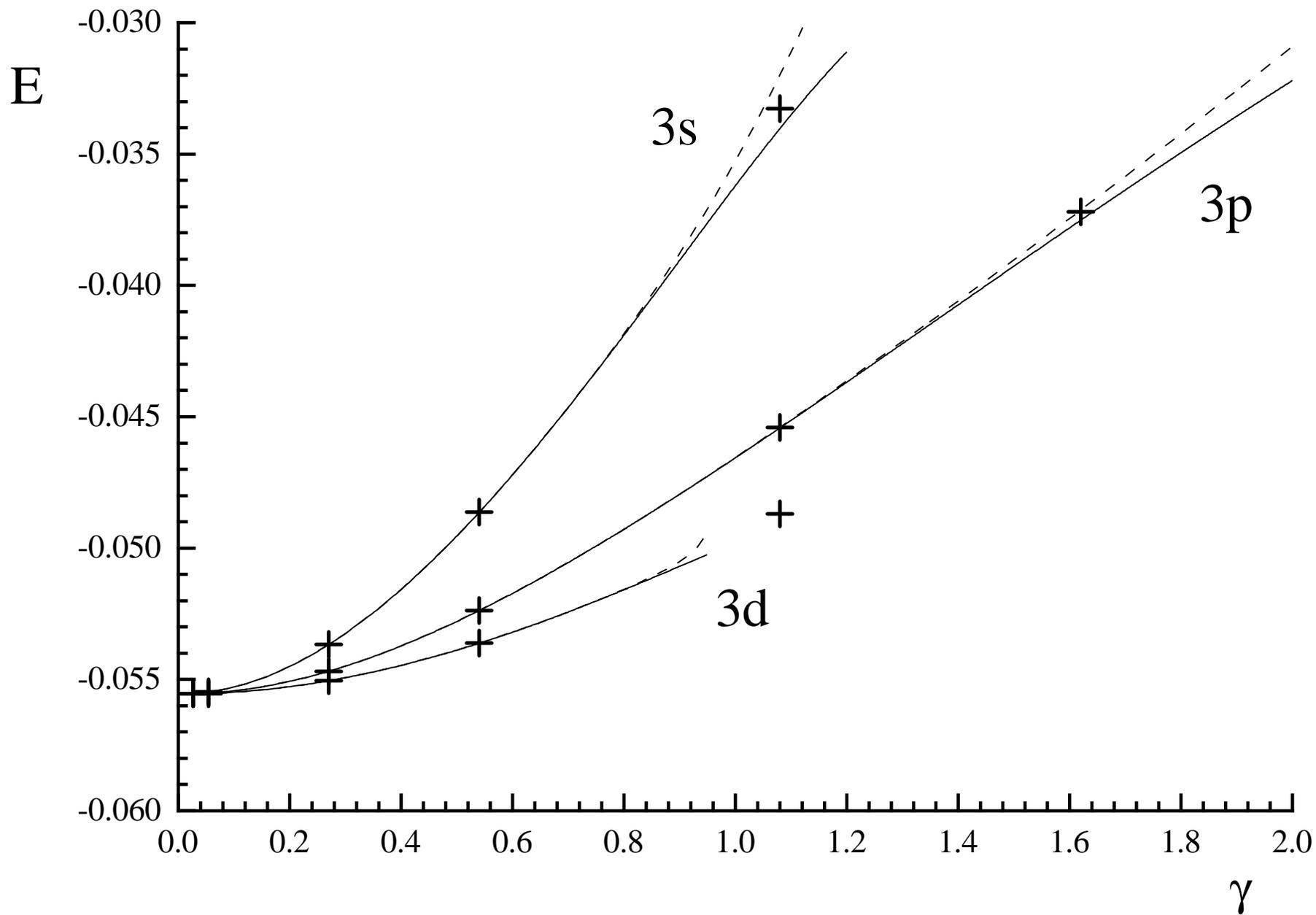
\end{document}